\documentclass[fleqn,usenatbib]{mnras}

\usepackage{aas_macros}
\usepackage{newtxtext,newtxmath}

\usepackage[T1]{fontenc}

\DeclareRobustCommand{\VAN}[3]{#2}
\let\VANthebibliography\thebibliography
\def\thebibliography{\DeclareRobustCommand{\VAN}[3]{##3}\VANthebibliography}

\usepackage{graphicx}	% Including figure files
\usepackage{amsmath}	% Advanced maths commands
\usepackage[section]{placeins} % Allows for the use of float barrier 
\usepackage[dvipsnames]{xcolor} %Allows use of more colours
\usepackage{aas_macros}
\newcommand{\drlength}{296\,007}	% DR11 total number of sources

\newcommand{\drbest}{202\,282} % DR11 best galaxies subset sources

\newcommand{\speczbest}{6558}

\newcommand{\chandralength}{868} % Chandra imaging total number of sources

\newcommand{\xrayagn}{710} % Number of xray sources in the UDS field and therefore number of xray AGN
\newcommand{\varxrayagn}{107} % Number of galaxies both variable and xray bright
\newcommand{\varagnall}{705} % Number of galaxies classed as variable using chi2 and therefore number of variable AGN
\newcommand{\varagn}{601} %Number of variable AGN we trust now 
\newcommand{\anyvarchandrasky}{292} %Number of variable AGN in the chandra imaging region

\newcommand{\semlc}{114\,243} % Number of sources that had light curves Lizzie sent me
\newcommand{\jbandmag}{22.66} %J-band magnitude limit for us to trust the variability amplitude of 0.2 or higher
\newcommand{\kbandmag}{23.25} %K-band magnitude limit for us to trust the variability amplitude of 0.2 or higher

\newcommand{\AGNoverlap}{37 per cent} %Percentage of variable AGN classed as X-ray bright when looking at the same area of sky

\newcommand{\textoverline}[1]{$\overline{\mbox{#1}}$}
\title[Increasing AGN completeness]{Increasing AGN sample completeness using long-term near-infrared variability}

\author[K.~Green et al.]{
K.~Green,$^{1}$\thanks{E-mail: karel.green@nottingham.ac.uk}
E.~Elmer,$^{1}$
D.~T.~Maltby,$^{1}$
O.~Almaini,$^{1}$
M.~Merrifield,$^{1}$
W.~G.~Hartley$^{2}$
\\
$^{1}$School of Physics and Astronomy, University of Nottingham, University Park, Nottingham, NG7 2RD, UK \\
$^{2}$Department of Astronomy, University of Geneva, CH-1205 Versoix, Switzerland 
}

\date{Accepted 2024 May 17. Received 2024 May 10; in original form 2024 March 7}

\pubyear{2023}

\begin{document}
\label{firstpage}
\pagerange{\pageref{firstpage}--\pageref{lastpage}}
\maketitle

\begin{abstract}
In this work, we use 8 years of deep near-infrared imaging to select and study a new set of \varagn~active galaxies identified through long-term near-infrared (NIR) variability in the UKIDSS Ultra Deep Survey (UDS). These objects are compared to \xrayagn~X-ray bright AGN detected by the {\em Chandra} X-ray observatory. We show that infrared variability and X-ray emission select distinct sets of active galaxies, finding only a \AGNoverlap~overlap of galaxies detected by both techniques and confirming NIR-variable AGN to be typically X-ray quiet. Examining the mass functions of the active galaxies shows that NIR variability detects AGN activity in galaxies over a significantly wider range of host stellar mass compared to X-ray detection. For example, at $z\sim 1$, 
variable AGN are identified among approximately $1$ per cent of galaxies in a roughly flat distribution above the stellar mass completeness limit ($>10^{9}\rm\,M_{\odot}$),  while X-ray detection primarily identifies AGN in galaxies of higher mass ($>10^{10}\rm\,M_{\odot}$). We conclude that long-term near-infrared variability provides an important new tool for obtaining more complete samples of AGN in deep survey fields.
\end{abstract}
\begin{keywords}
surveys -- galaxies: active -- galaxies: luminosity function, mass function -- infrared: galaxies -- X-rays: galaxies
\end{keywords}

%%%%%%%%%%%%%%%%%%%%%%%%%%%%%%%%%%%%%%%%%%%%%%%%%%

%%%%%%%%%%%%%%%%% BODY OF PAPER %%%%%%%%%%%%%%%%%%

\section{Introduction}
Active Galactic Nuclei (AGN)  have long been observed within the Universe \citep{seyfert_nuclear_1943} and are characterised by their dynamic multi-wavelength emission that spans the entire electromagnetic spectrum (e.g., \citealt{edelson_far-infrared_1987}; \citealt{collier_characteristic_2001}; \citealt{cackett_reverberation_2021}). This AGN emission, which can often outshine the entire stellar output of the host galaxy, is believed to arise from accretion of material onto a central supermassive black hole  (\citealt{salpeter_accretion_1964}; \citealt{kormendy_inward_1995}). AGN emission is invoked across many aspects of galaxy evolution to help explain observed phenomena, but this activity, and the underlying mechanisms that drive it, is still not fully understood.

Due to the small  physical size of the AGN's central engine, accretion and emission from AGN occurs on scales many orders of magnitude smaller than that of the host galaxy. Consequently, resolving the central nucleus is simply not feasible for galaxies outside our local neighbourhood (e.g., \citealt{padovani_active_2017}). The difficulty in unpicking AGN emission from that of its host galaxy also poses a significant barrier in understanding the structures present within an AGN, and the impacts of the accretion process on the surrounding region (\citealt{grogin_agn_2005}; \citealt{gabor_active_2009}; \citealt{pierce_effects_2010}; \citealt{fan_structure_2014}). As a result, despite many theories  within the literature, there is no firm consensus on the potential trigger (or triggers) that activates an AGN. Furthermore, we are currently unable to observe any given galaxy and know with certainty how and why it is, has been, or will become, active in its lifetime. In order to determine robust answers for these questions, a complete census of active galaxies in the Universe is necessary to allow for the study of the cosmic evolution of AGN.

Due to the nature of the accretion process, AGN emission is not static, but varies in flux over time at every wavelength it has been observed (e.g., \citealt{fitch_light_1967}, \citealt{sanchez_near-infrared_2017}).  It has been hypothesised that much of this variability is caused by stochastic processes in the accretion disc, and as such allows for the study of AGN on the relatively tiny scales they span compared to the size of a galaxy. Variability timescales have also been found to increase with wavelength. For example, emission at higher frequencies, such as X-rays, vary within hours, while UV and optical variations take days to weeks. The longer wavelengths, such as infrared (IR), typically vary on timescales of months or years (e.g., \citealt{berk_ensemble_2004}). AGN emission theory couples the wavelengths of emission to structures within the AGN itself, with longer wavelengths being generated at larger distances from the core. For example, X-rays are believed to be generated in the X-ray corona, UV and optical emission in the accretion disk, and IR emission originating in a surrounding dusty torus (e.g., \citealt{bianchi_active_2022}).

Variability has been shown to be a useful tool in selecting AGN, being able to complement other selection methods by succeeding where they fail. X-ray emission, for example, is commonly thought to have a high probability of selecting the most complete sample of active galaxies (e.g., \citealt{suh_multi-wavelength_2019}, \citealt{pouliasis_robust_2019}), but this method is intrinsically biased against X-ray-faint objects. In \cite{trevese_variability-selected_2008}, a study of low-luminosity AGN found that only 44 per cent of their optical-variability--selected active galaxies have associated X-ray emission. \cite{pouliasis_robust_2019} also selected AGN via optical variability and X-ray emission, but additionally employed the use of IR colour selection. They find that 23 per cent of their variable sample have associated X-ray emission, and only 12 per cent of their variable sample satisfies the IR colour selection criteria for the presence of an AGN. More recently, \cite{lyu_agn_2022} used a wide combination of selection techniques, including radio emission, X-ray detection, variability, UV--IR SED analysis, optical spectroscopy, and IR colour selection in an attempt to find a complete sample of active galaxies in the GOODS-S field. They concluded that no single method was able to select a complete sample of galaxies, and the overlap in the AGN found by different techniques was small.

AGN variability is a relatively unexplored property at infrared wavelengths due to the time required to observe it, but one which is believed to provide significant insight into the nature of the outer edges of the AGN system. In principle, this technique may also provide an advantage in identifying heavily-obscured and/or high-redshift AGN, which may be missed by optical/UV selection techniques.
Furthermore, deep wide-field imaging can allow us to study thousands of galaxies simultaneously. 

Until recently, studies of IR variability either focused on a handful of individual objects (\citealt{cutri_variability_1985}, \citealt{lira_optical_2011}, \citealt{lira_long-term_2015}),  observed AGN selected by other means (\citealt{kouzuma_ensemble_2012}, \citealt{sanchez_near-infrared_2017}, \citealt{son_mid-infrared_2022}), or studied light curves that are not well sampled \citep{edelson_far-infrared_1987} or constructed from observations from multiple different surveys \citep{neugebauer_near-infrared_1989}. However, \cite{elmer_long-term_2020} were the first to select AGN in large numbers based purely on their NIR variability using a single deep data set, and it is this sample we build upon in this work.

In this paper, we investigate the properties of active galaxies selected using NIR variability. We use $J$-band light curves to select significantly variable AGN in addition to the $K$-band light curves and method described in \citealt{elmer_long-term_2020}. We then compare the properties of these galaxies to a sample of AGN selected via X-ray selection,  to determine if and how the properties of active galaxies differ depending on the detection method. We make use of nearly a decade of IR observations provided by the Ultra Deep Survey (UDS; Almaini et al., in prep); the deepest component of the UKIRT (United Kingdom Infra-red Telescope) Infrared Deep Sky Survey \citep[UKIDSS;][]{lawrence_ukirt_2007}. We use this data to select~\varagn~active galaxies between 0 < z $\leq$ 5 based purely on their IR variability using the technique developed, as well as the data constructed, in \citet{elmer_long-term_2020}. We then compare the properties of these active galaxies to those found using X-ray emission from the X-UDS survey \citep{kocevski_x-uds_2018}, as well as a control sample of inactive galaxies.

The layout of this paper is as follows. In Section \ref{sec:data} we describe the data used in this research, and in Section \ref{sec: method} we describe the methodology used to select and study the active galaxies. Section \ref{sec:results} contains the results, and finally Section \ref{sec:conclusion} provides a summary and the conclusions we draw from this analysis. All magnitudes stated are AB magnitudes. We use a $\Lambda$ cold dark matter ($\Lambda$CDM) cosmology, with $H_{0}$ = 70\,kms$^{-1}$Mpc$^{-1}$, $\Omega_{\Lambda} = 0.7$ and $\Omega_{\rm m} = 0.3$.

\vspace{-0.5cm}

\section{Data}
\label{sec:data}

\begin{figure}
    \centering
    \includegraphics[width=1\linewidth]{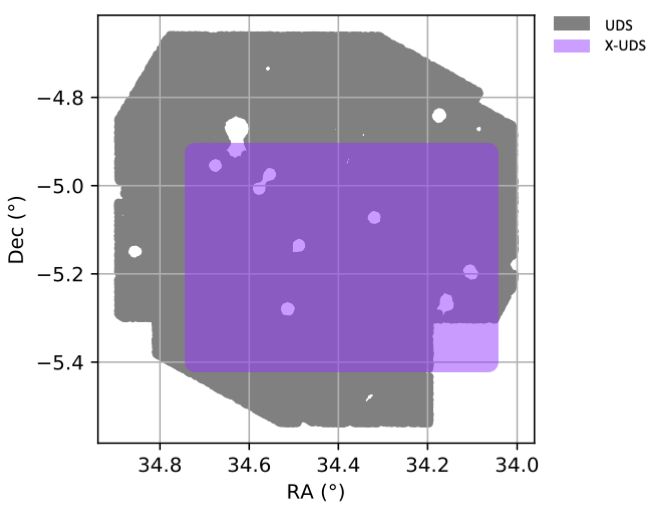}
    \vspace{-0.25cm}
    \caption{The survey region of the Ultra Deep Survey (UDS) and the {\em Chandra} Legacy Survey of the UDS field (X-UDS). The grey shaded area shows the subset of the ground-based UDS field where there is reliable photometry in all 12 photometric bands, and the purple shaded area shows the coverage from the {\em Chandra} X-UDS imaging.}
    \label{fig:Data_field}
    %/data/Var/Code/imaging_area.py
\end{figure}

In this work, we make use of data from two deep surveys: the UKIDSS Ultra Deep Survey (UDS; Almaini et al., in prep); combined with the {\em Chandra} Legacy Survey of the UDS field (X-UDS; \citealt{kocevski_x-uds_2018}). A plot of the extent of the fields is shown in Figure \ref{fig:Data_field}.

\vspace{-0.25cm}

\subsection{UDS}
\label{subsec: uds dr11}
The UDS is the deepest of the UKIDSS surveys and the deepest NIR survey over $\sim1\rm\,deg^{2}$. Its $JHK$ band imaging was taken over an 8 year period from 2005--2013 using the WFCAM instrument at UKIRT \citep{casali_ukirt_2007}. Although not primarily designed to study variability, the collection of the data spread over eight years offers a powerful resource for long-term infrared variability studies. In addition to this, the field is well studied and as such has additional imaging in many other wavebands including:

%\vspace{1cm}

\begin{itemize}
    \item $u$'-band imaging from the Canada--France--Hawaii Telescope (CFHT) MegaCam. 
    \item     $B$, $V$, $R$, $i$' and $z$'-band optical imaging from the Subaru {\em XMM-Newton} Deep Survey \citep[SXDS; ][]{furusawa_subaruxmm-newton_2008}.
    \item $Y$-band imaging from the VISTA VIDEO survey \citep{jarvis_vista_2013}.
    \item 3.6\,$\mu$m and 4.4\,$\mu$m mid-infrared (MIR) IRAC imaging from the {\em Spitzer} UDS Legacy Programme (PI: Dunlop).
\end{itemize}

The DR11 is the latest data release from the Ultra Deep Survey (UDS).
This data release has \drlength~$K$-band detected sources and  5$\sigma$ limiting depths of $J$ = 25.6, $H$ = 24.8 and $K$ = 25.3 (AB in 2-arcsec diameter apertures; \citealt{almaini_massive_2017}; Almaini et al., in prep). From this data release we make use of the `\textit{best galaxies}' subset which covers 0.62$\rm\,deg^{2}$, and comprises objects identified in unmasked regions of the field with reliable photometry in all 12 bands, the deepest IRAC imaging, and with galactic stars removed. Masked regions correspond to boundaries of the science image, artefacts, bright stars and detector cross-talk. The \textit{best galaxies} subset contains \drbest~$K$-band detections.

\begin{figure}
    \centering
    \includegraphics[width=\columnwidth]{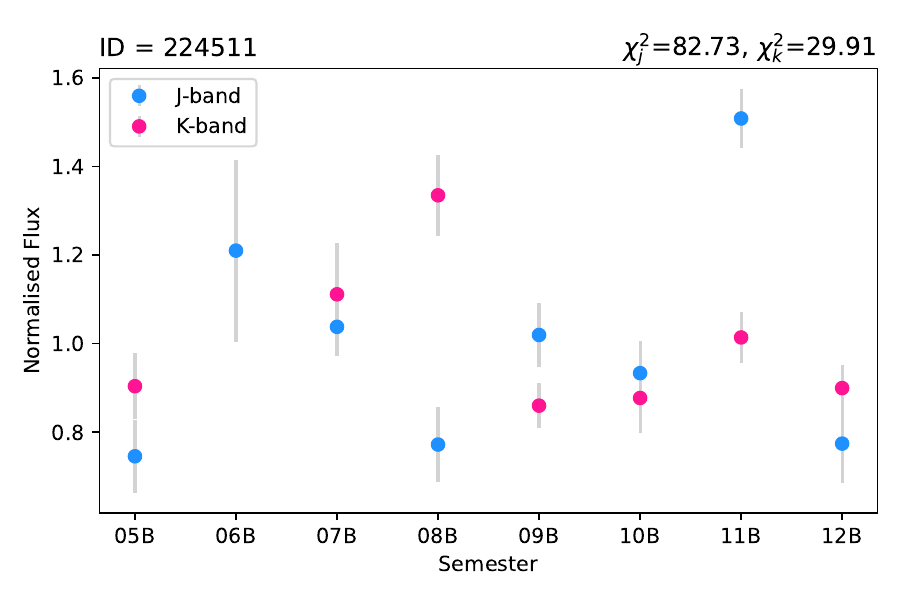}
    \includegraphics[width=\columnwidth]{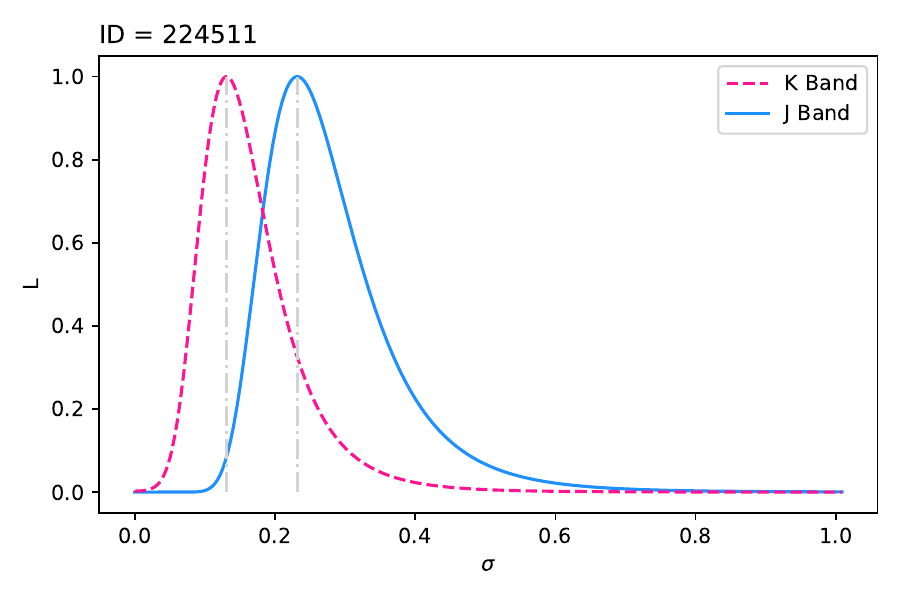}
    \vspace{-0.30cm}
    \caption{Example light curves for a variable galaxy candidate in both the $J$ and $K$-band (top panel) and the corresponding normalised likelihood curves (bottom panel). The threshold for recovery of AGN variability from a given light curve is $\chi^{2}_{K}>30$ and $\chi^{2}_{J}$ > 32.08. As shown by the grey lines in the bottom panel, the variability amplitude is determined by selecting the corresponding value for $\sigma_{\rm Q}$ that maximises the likelihood function. This galaxy is only classified as formally variable in the $J$-band (blue), as shown by the value of the $\chi^{2}$.}
    \label{fig:max_like_ex}
\end{figure}

\subsubsection{Redshifts}
Of the \drbest~galaxies, \speczbest~have secure spectroscopic redshifts provided by complementary spectroscopic surveys \citep[e.g., UDSz, VANDELS;][]{bradshaw_high-velocity_2013, mclure_sizes_2013, maltby_identification_2016, mclure_vandels_2018, pentericci_vandels_2018}. For non-variable objects, spectroscopic redshifts are used (where available); otherwise, photometric redshifts are adopted.

Photometric redshifts for non-variable objects were calculated according to the methodology described in \citet{simpson_prevalence_2013}.  For this, the 12-band photometry was fit using a grid of galaxy templates constructed from the stellar population synthesis models of \citet{bruzual_stellar_2003}, and used the publicly available code {\sc eazy} \citep{brammer_eazy_2008}. These photometric redshifts have a typical accuracy of $\frac{\Delta z}{(1+z)} \approx 0.018$.  Additional properties used in this work (e.g.\ stellar mass and luminosity; Almaini et al., in prep), are based on these redshifts and are entirely adequate for our purposes.

In this paper, we identify IR-variable objects (see Section \ref{sec: selecting var AGN}). For these objects, spectroscopic redshifts were used (where possible) over photometric determinations. For the variability detected sources for which spectroscopic redshifts were not available, inaccuracies in the photometric redshifts are a potential concern due to the AGN emission not being accounted for. To address this, we have recalculated the photometric redshifts for these objects via the {\sc eazy} redshift fitting code. Here additional reddened and un-reddened AGN templates were included in the fitting. Comparing the median absolute deviation of $\frac{\Delta z}{(1+z)}$ before and after including the recalculated photometric redshifts, indicates the redshifts improve only marginally from $\sigma_{\rm MAD}=0.10$ to $\sigma_{\rm MAD}=0.098$.  This demonstrates the initial photometric redshift values to be largely robust. Regardless, the recalculated redshifts for the IR-variable active galaxies are used throughout this work, though we note that using the original photometric redshifts made no significant difference to our analysis or conclusions.

\subsubsection{Stellar Masses}
Stellar masses (Almaini et al., in prep), were calculated using the method described in \cite{simpson_prevalence_2013}.  Here, the 12-band photometry was fit to a grid of synthetic SEDs from the stellar population models of \cite{bruzual_stellar_2003} with a  \cite{chabrier_galactic_2003} initial mass function. Typical uncertainties in these stellar masses are of the order $\pm$0.1 dex. The redshift-dependent, 90 per cent stellar mass completeness limits  were calculated using the method described in \citet{pozzetti_zcosmos_2010}. A second order polynomial was fit to the resulting completeness limits across a wide redshift range to give a $90$ per cent mass completeness curve of the form ${\rm log_{10}}(M_{*}/{\rm M_{\odot}}$) = -0.04$z^{2}$ + 0.67$z$ + 8.13.

\begin{figure}
    \centering
    \includegraphics[width=\columnwidth]{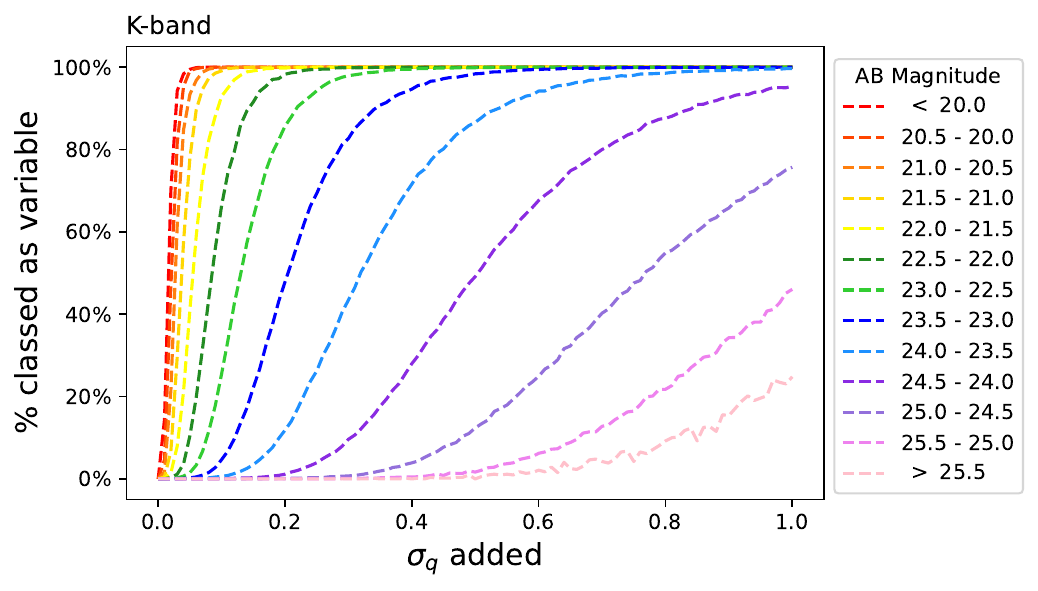}
    \includegraphics[width=\columnwidth]{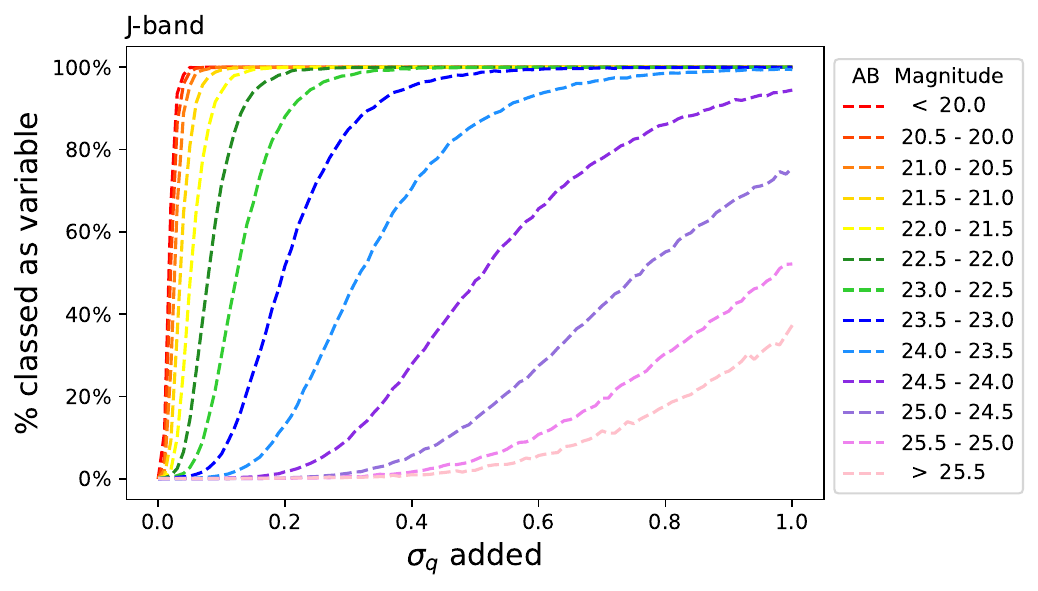}
    \caption{The fraction of galaxies classed as hosting variable AGN compared to the level of artificial variability added to the inactive galaxy light curves.  Results are shown for the $K$-band (top panel) and $J$-band (bottom panel). Each dashed line represents a different apparent magnitude bin. In the brightest bins, lower uncertainties on measurements allow for low amplitudes of variability to be recovered. However, larger variability amplitudes are required for AGN to be detected in the faintest galaxies.}
    \label{fig:var_pcnt_curve}
\end{figure}

\subsection{X-UDS}
\label{xuds}
The X-UDS survey is a deep X-ray survey of a sub-region of the UDS field carried out by the {\em Chandra} X-ray Observatory's Advanced CCD Imaging Spectrometer (ACIS).  It comprises a mosaic of 25 observations of 50\,ks exposure, totalling 1.25\,Ms of imaging. It covers 0.33\,deg$^{2}$ in total \citep{kocevski_x-uds_2018} and has a flux limit of 4.4$\times$10$^{-16}$\,erg\,s$^{-1}$cm$^{-2}$ in the full band (0.5--10\,keV). In total, \chandralength~point sources were identified.

This data set was matched to the UDS DR11 $K$-band catalogue using the methodology described in \citet{civano_chandra_2012}, which adopted the maximum likelihood method described in \citet{sutherland_likelihood_1992}.  Through this analysis, we find that \xrayagn~of these point sources are associated with $K$-band--selected galaxies in the UDS DR11.

\begin{figure}
    \centering
    \includegraphics[width=\columnwidth]{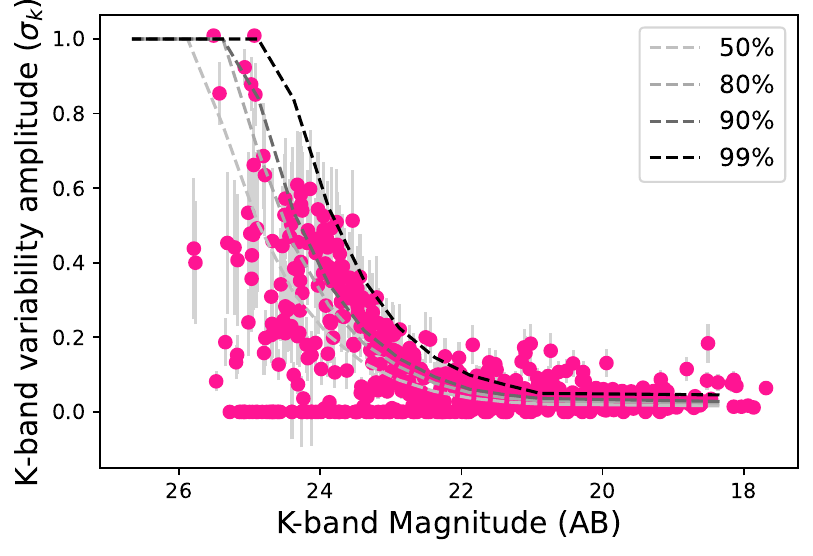}
    \includegraphics[width=\columnwidth]{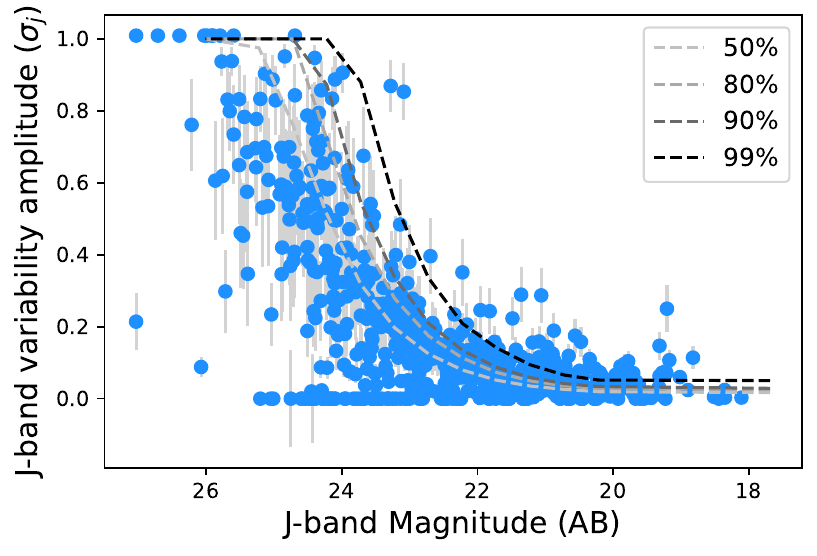}
    \caption{AB magnitude vs. variability amplitude for the \textit{real} variable active galaxies. Pink points indicate galaxies selected as hosting variable AGN in the $K$-band (top panel), while blue points show the same data but for the $J$-band (bottom panel). Grey 1$\sigma$ errors are based on assuming a Gaussian fit to the likelihood curves. Overlaid are detection limit curves for different levels of completeness. We use the 90 per cent curve as this threshold provides a high level of  completeness without over-reducing the sample size.}
    \label{fig:detect_lim_curve}
\end{figure}

\section{Method}
\label{sec: method}

\subsection{Selecting AGN based on their IR variability}
\label{sec: selecting var AGN}
IR-variable AGN are selected from the UDS DR11 using the technique first developed and used in \cite{elmer_long-term_2020}. Here, the authors were able to construct NIR light curves from multiple flux measurements that spanned 7 semesters in the $K$-band. After the publication of \citealt{elmer_long-term_2020}, additional $J$-band imaging was incorporated into the dataset. First an updated catalogue of galactic stars was developed for the UDS DR11 release, none of which were included in the initial $K$-band sample selection. These stars were used to improve the PSF images prior to convolving the data to match the semester with the poorest PSF (see Section 2.2 of \citealt{elmer_long-term_2020}), allowing for an improved sample selection in the $K$-band. Furthermore, the 2006B semester in the $J$-band had sufficient observations to allow it to be included in the light curves. 
This gives the $J$-band a total of 8 epochs for study.

\begin{figure}
    \centering
    \includegraphics[width=\columnwidth]{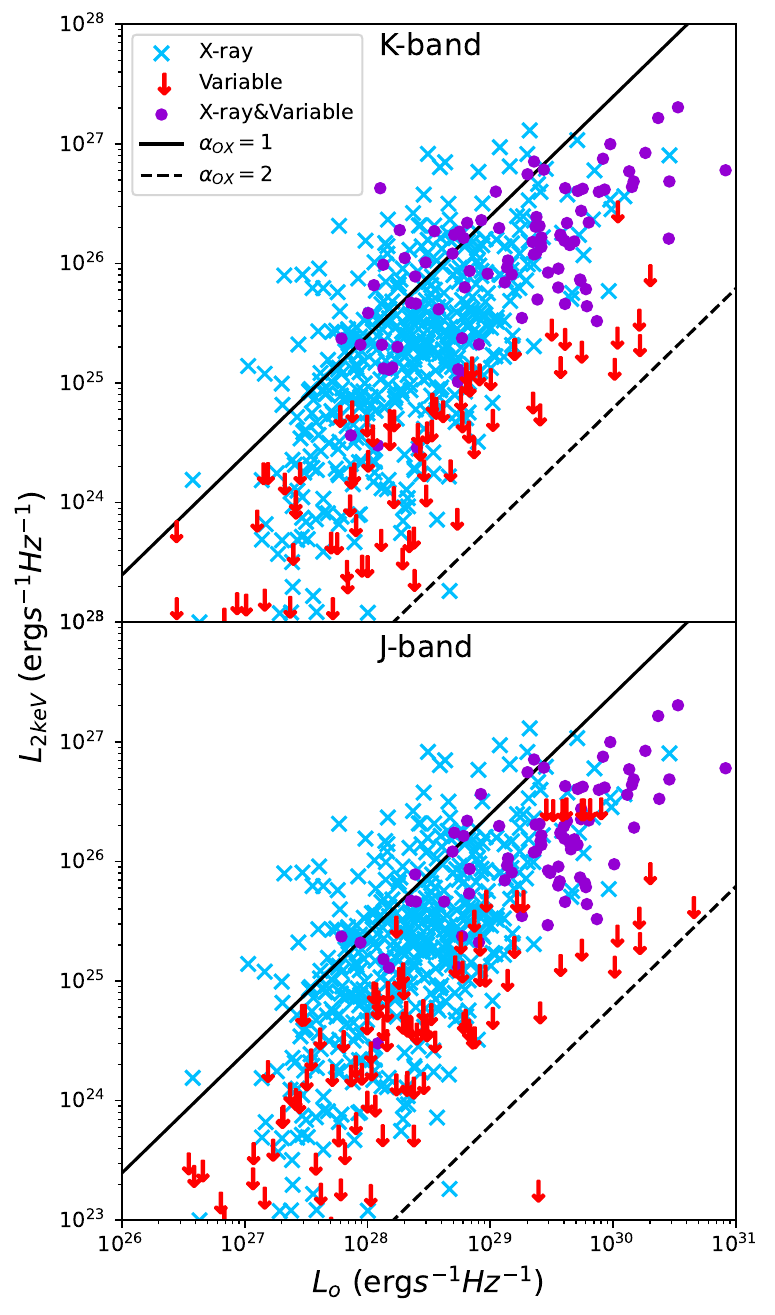}
    \caption{Monochromatic 2\,keV X-ray luminosity vs. rest-frame optical luminosity at 2500$\,\text{\r{AA}}$ for objects imaged within the {\em Chandra} region of the UDS field. X-ray bright active galaxies are denoted by blue crosses and AGN detected by both NIR variability and X-ray emission are highlighted as purple dots. Upper limits are used for variability-detected active galaxies without X-ray detections, and are shown by red arrows. The top plot shows $K$-band detected active galaxies and the bottom plot shows $J$-band detected active galaxies. Black lines indicate the slopes corresponding to a spectral index ($\alpha_{\rm ox}$) of $\alpha_{\rm ox}$=1 (solid) and $\alpha_{\rm ox}$=2 (dashed), where $\alpha_{\rm ox}$ is the spectral index between the optical and the X-ray.}
    \label{fig:lx_lo}
\end{figure}

We further select only objects with reliable (unmasked) photometry in all 12 bands, to ensure the most reliable photometric redshifts. Objects identified as galactic stars were excluded, as were any galaxies close to cross-talk detector artefacts. We also exclude any objects with formally negative flux values in any individual semester, as described in \cite{elmer_long-term_2020}. This final step largely excluded very faint galaxies at $K>25$, close to or beyond the formal detection limit of the survey. Inspecting this faint subgroup reveals that all would be excluded by the eventual $J$-band magnitude cut described in 
Section \ref{sec: var detect lim}, so have no impact on our analysis. Overall, these rejection criteria yield a final sample of $114\,243$ light curves for this study.

Errors on the flux measurements were self-calibrated, based on the population variance in flux of the non-variable objects at the same apparent magnitude, as this provided a better fit to the data than the errors generated by \textsc{SExtractor} (see Figure 3 in \citealt{elmer_long-term_2020}). A full description of this process can be found in Section 2.2 of \cite{elmer_long-term_2020}.  

IR-variable AGN were selected using a $\chi^{2}$ analysis applied to the IR light curves based on the null hypothesis that flux measured from the object is constant with time. Thus we calculate
\begin{equation}
    \chi^{2} = \sum_{i}\frac{(F_{i}-\bar{F})^{2}}{\sigma_{i}^{2}},
    \label{eqtn:chi2}
\end{equation}
where for each semester $i$, $F_{i}$ is the flux of an object, $\sigma_{i}$ is the corresponding uncertainty and $\bar{F}$ is the mean flux of that object across all imaging epochs. A threshold of $\chi^{2} > 30$ was used for the $K$-band and we use $\chi^{2} > 32.08$ for the $J$-band. The differing $\chi^{2}$ limits give the same p-value for both $\chi^{2}$ distributions based on the $J$ and $K$-band light curves, accounting for the different number of imaging epochs and therefore differing degrees of freedom. If a galaxy satisfies either or both conditions, it is classed as hosting a variable AGN.

Utilising this method to select significantly variable objects via the $\chi^{2}$ analysis, we are able to identify a sample of \varagnall~candidate active galaxies for study (430 from $K$-band variability, 471 from $J$-band variability, with 196 detected by both). We recover all of the 393 $K$-band selected AGN found in \citealt{elmer_long-term_2020}, with a modest increase  due to the improvements in the data analysis  since the publication of that work. From the initial data set of \semlc~galaxies with IR light curves, we would expect a total of 4 false positives based on a $\chi^{2}$ distribution with this number of degrees of freedom. 

To search for potential supernovae in the sample, tests were carried out according to Section 5 of \citealt{elmer_long-term_2020}. We identified 36 
of the 471 $J$-band variables (7.6 per cent) for which the variability is largely driven by one epoch.
Inspection of both the $J$ and $K$-band imaging and  comparing epochs revealed no obvious signs of supernovae (e.g., off-nuclear sources). 
No additional supernova candidates were identified among the new $K$-band variables not analysed by \citealt{elmer_long-term_2020}. 
Removing this small number of galaxies had no influence on any of the key conclusions presented in this work.

\begin{figure}
    \centering
    \includegraphics[width=\linewidth]{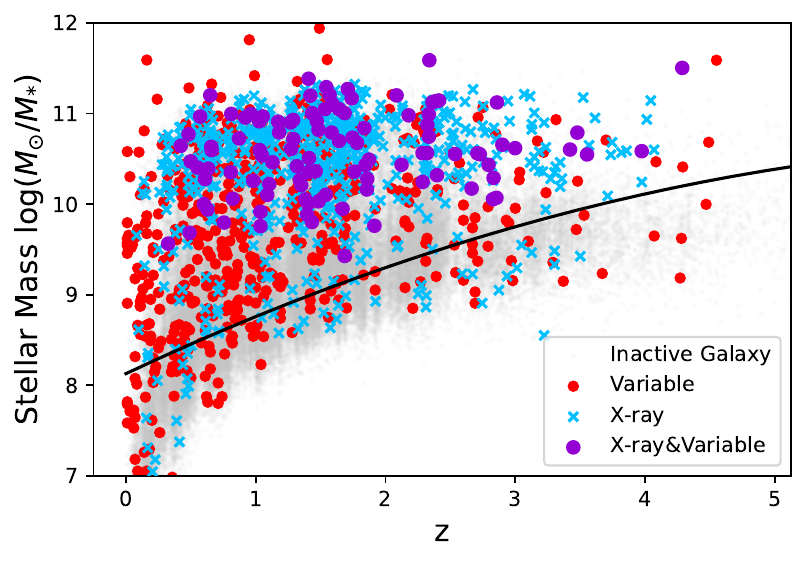}
    \caption{Stellar mass vs. redshift distribution for the galaxies in this study. X-ray bright active galaxies are shown as blue crosses, red circles denote variability-detected active galaxies, and purple circles show dual-detected active galaxies. Grey points are inactive galaxies, and the black curve shows the 90 per cent stellar mass completeness limit as a function of redshift.}
    \label{fig:mass_vs_z}
\end{figure}

\subsection{Galaxies on the edge of the science image}
Visual inspection revealed a suspicious sample of 104 variable objects  located very close to the detector edges.
These objects were inspected individually within each epoch, and approximately $\sim$33 per cent were identified as potentially fake variables. All objects were found to be real galaxies, but in many cases false variability could be attributed to problems with background subtraction at the detector edges. As such, we re-ran all analysis both with and without this sub-sample of galaxies, and found no significant change in the overall reported trends. Therefore, we removed these edge objects entirely from the analysis to ensure robust results. The remaining sample consists of \varagn~galaxies.
We note that the pattern of false variables occurred only at the outer edges of the UDS imaging. There was no evidence for false positives close to the internal WFCAM chip boundaries in the UDS mosaic.

\begin{figure}
    \centering
    \includegraphics[width=\linewidth]{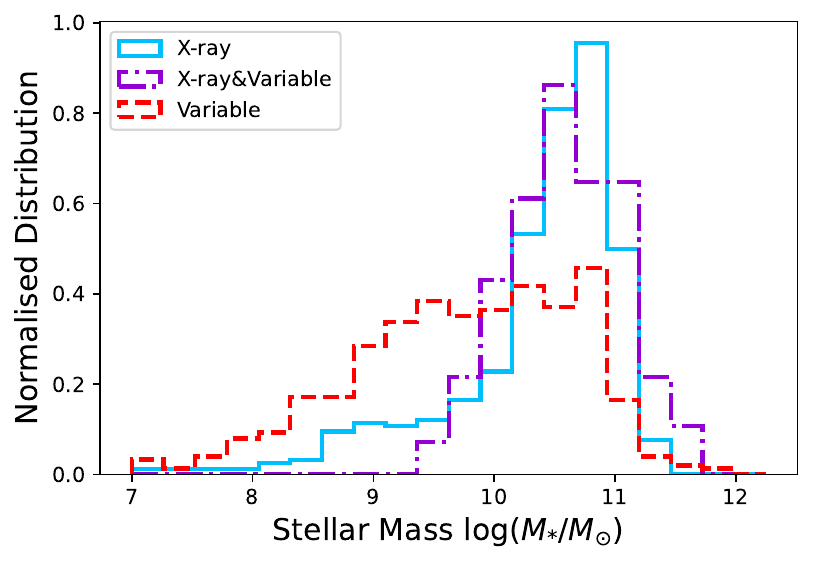}
    \caption{Normalised histograms showing the distribution in stellar mass of active galaxies in the sample. Galaxies are separated according to detection method, where the blue solid-lined histogram represents X-ray selected active galaxies, the red-dashed histogram shows objects classified as hosting variable AGN only, and the purple-dotted histogram shows objects that were detected using both methods. A Kolmogorov-Smirnov test confirms that the variability detected sample has a significantly different distribution in stellar masses compared to the X-ray population, rejecting the null hypothesis that they are drawn from the same underlying mass distribution with a significance of $>99.99$ per cent.}
    \label{fig:mass_hist}
\end{figure}

\begin{figure*}
    \centering
    \includegraphics[width=0.85\textwidth]{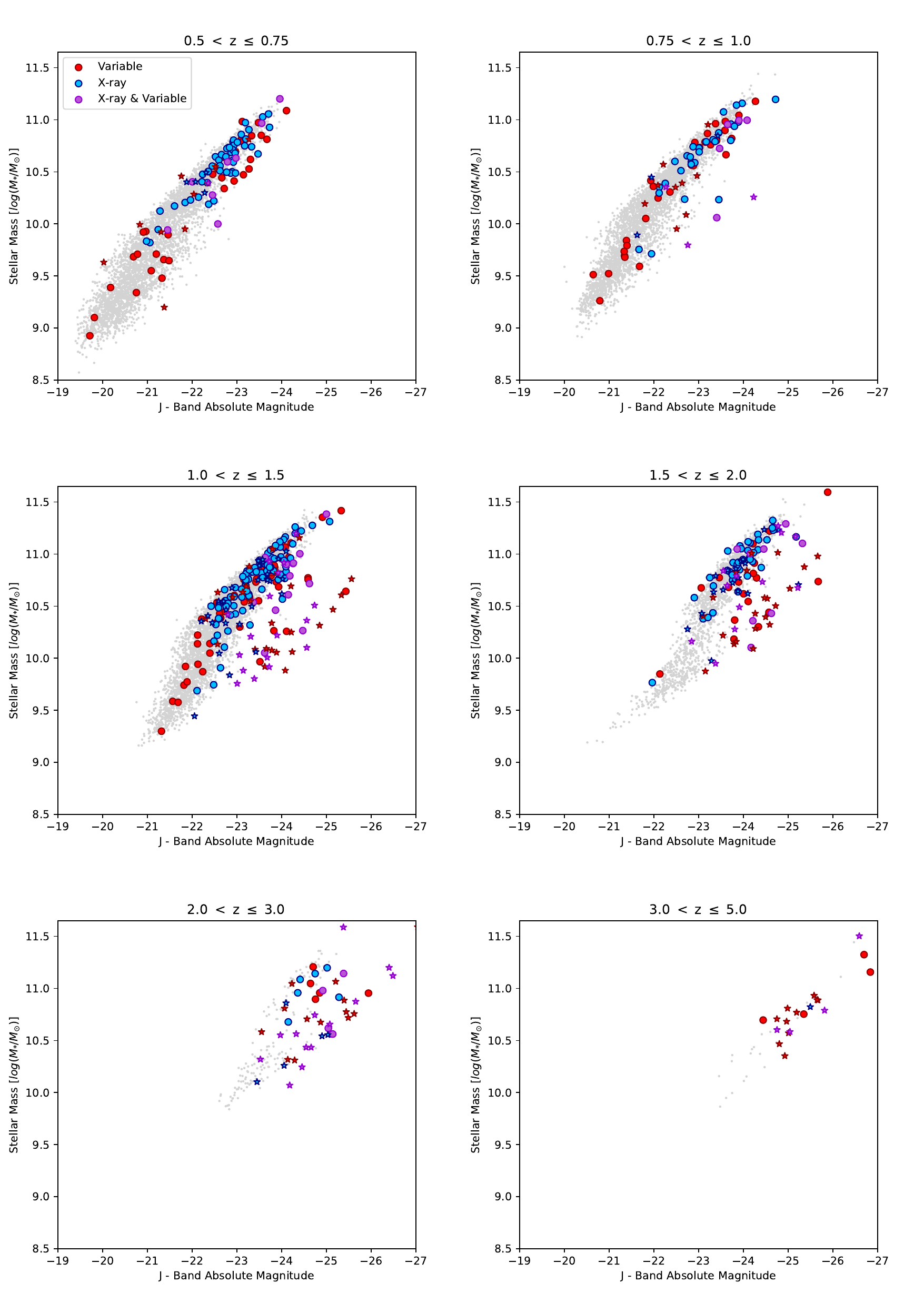}
    \caption{Stellar mass vs. $J$-band absolute magnitude for X-ray detected active galaxies (blue points), variability-detected active galaxies (red points) and dual-detected active galaxies (purple points) as a function of redshift. Inactive galaxies are plotted as grey points and all galaxies lie above the 90 per cent mass completeness limit. Corresponding coloured stars indicate an active galaxy that has a stellarity index of $\geq$ 90\% according to the \textsc{class star} parameter from \textsc{SExtractor}.}
    \label{fig:mass_mag}
\end{figure*}

\subsection{Measuring real variability: maximum likelihood}
\label{sec: max likely}
For any measurement of an active galaxy light curve, we know that the observed dispersion in flux ($\sigma_{\rm obs}$) is due to a combination of the intrinsic magnitude variations due to the active galactic nucleus ($\sigma_{\rm Q}$) and the measurement noise ($\sigma_{\rm noise}$) such that 
\begin{equation}
    \sigma^{2}_{\rm obs} = \sigma^{2}_{\rm Q} + \sigma^{2}_{\rm noise}.
\end{equation}
Consequently, if we are to measure the properties of the AGN, we have to be able to separate out these two effects.

As such, to calculate the fractional variability amplitude due to AGN emission ($\sigma_{\rm Q}$), we make use of a maximum likelihood technique. This method, developed in \citet{almaini_X-ray_2000}, allows for an estimate of the noise-subtracted intrinsic fractional variance ($\sigma_{\rm Q}$) of the AGN, within the observed variability ($\sigma_{\rm obs}$). We first take any given AGN light curve and generate a likelihood function 
\begin{equation}
    L(\sigma_{\rm Q}\,|\,x_{i},\sigma_{i}) = {\prod_{i=1}^{N} \frac{{\rm exp}[{-(x_{i}-\bar{x})^{2}}/2(\sigma_{i}^{2}+\sigma_{\rm Q}^{2})]}{(2\pi)^{1/2}(\sigma_{i}^{2}+\sigma_{\rm Q}^{2})^{1/2}}}.
    \label{eq:max_like}
\end{equation}
To generate each point in the likelihood curve, we normalise each galaxy light curve such that the average flux ($\bar{x}$) is unity. We then step through possible values of fractional AGN variability ($\sigma_{\rm Q}$) using the measurement of the flux in each semester ($x_{i}$) and its corresponding measurement error ($\sigma_{i}$). To get the final likelihood measurement, we take the product of the N values that have been calculated where N is the number of flux measurements in a given light curve.

In order to determine the most likely value of $\sigma_{\rm Q}$ for each active galaxy, we simply select the value of $\sigma_{\rm Q}$ that maximises the likelihood function, as seen in Figure \ref{fig:max_like_ex}. Here the 1$\sigma$ errors are calculated by assuming the likelihood curve has a Gaussian shape.

\begin{figure}
    \centering
    \includegraphics[width=0.9\linewidth]{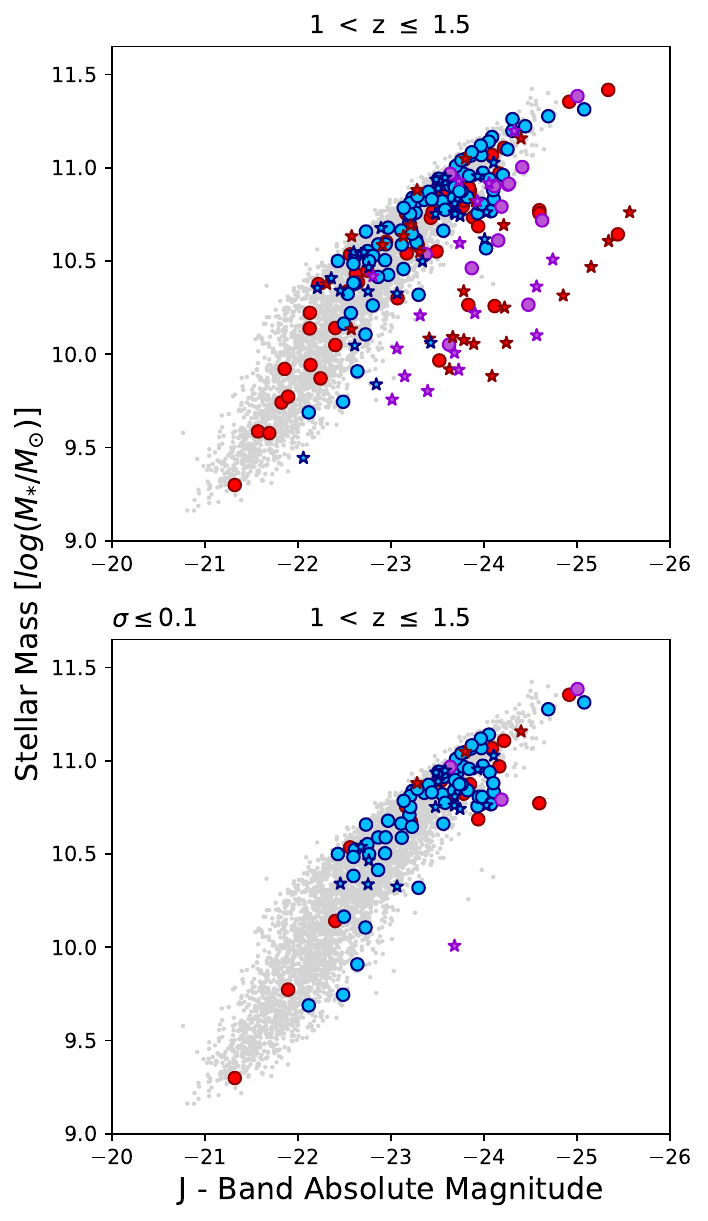}
    \caption{Stellar mass vs.~$J$-band absolute magnitude for galaxies between $1 < \ z \leq 1.5$. Plotted are active galaxies (AGN) that are either X-ray-detected (blue points), variability-detected (red points), or dual-detected (purple points), as well as a comparison sample of inactive galaxies (grey points). Correspondingly-coloured stars indicate an active galaxy that has a stellarity index of $\geq$ 90\% according to the \textsc{class star} parameter from \textsc{SExtractor}. The bottom plot is as the top, but only shows active galaxies with (intrinsic) variability amplitudes of $\sigma_{J} \leq 0.1$. All galaxies plotted lie above the 90\% mass completeness limit.}
    \label{fig:mass_mag_lowsig}
\end{figure}

\subsection{Variability detection limit}
\label{sec: var detect lim}

The $\chi^{2}$ detection method is inevitably sensitive to photometric errors, and thus in fainter galaxies we can only expect to detect variability at higher fractional amplitude.  We can attempt to quantify this effect, to estimate our completeness in identifying variable AGN with a range of intrinsic variability amplitudes ($\sigma_{\rm Q}$), and over a range of apparent magnitudes. To begin, we remove any active galaxies via their IR variability or X-ray emission from the initial sample of \semlc~galaxies and then split the remaining inactive galaxies into bins based on apparent magnitude. Then, to simulate variability in the inactive galaxies, we artificially introduce variability to their light curves assuming a Gaussian distribution of variations. We require this (variable) component to have a mean of zero and a standard deviation equal to the degree of variability we are simulating, and add this distribution to the normalised inactive galaxies light curves. We then re-run the $\chi^{2}$ and maximum likelihood analysis on the `new' light curve for each galaxy, which now contains simulated variability. This allows us to determine if it is now classed as hosting a significantly variable object by the $\chi^{2}$ analysis and to ascertain the most likely amplitude  of simulated variability using the maximum likelihood technique. This procedure is repeated for a range of variability amplitudes and for each apparent magnitude bin. The results are then used to determine, for a given apparent magnitude, the minimum fractional variability amplitude required for an AGN to be consistently detectable. This also allows us to compare the known simulated variability amplitude to the amplitude recovered via the maximum likelihood method. 

We find that the maximum likelihood technique accurately recovers the amplitude of input AGN variability. Sampling galaxies from each magnitude bin, the average difference in variability amplitude is consistently of the order of 1 per cent. Using these simulations, we are also able to calculate the fraction of galaxies classed as hosting an AGN for every value of simulated $\sigma_{\rm Q}$ in each magnitude bin (see Figure \ref{fig:var_pcnt_curve}). This provides a suitable estimate for the completeness of our NIR variables, as a function of both apparent magnitude and variability amplitude.

As seen in Figure \ref{fig:detect_lim_curve}, for a given value of an AGN's fractional variability amplitude and a selected completeness level, we can determine the apparent magnitude required for that selected variability amplitude to be consistently detected to the desired completeness. Throughout this work, we adopt the 90 per cent completeness threshold for a variability amplitude of 20 per cent or higher. From the $J$-band curve in Figure \ref{fig:detect_lim_curve}, we find this requires our galaxies to have apparent magnitudes of $m_{J} \leq$ \jbandmag. In the $K$-band, the same limit requires an apparent magnitude of $m_{K} \leq$ \kbandmag. We use the $J$-band limit in this work as it is stricter, and removes all galaxies excluded by the $K$-band limit.

\section{Results}
\label{sec:results}

\subsection{X-ray faint active galaxy populations}
\label{sec:x-ray faint AGN}

We applied the AGN variability detection method described in Section \ref{sec: method} to the UDS DR11, and compared the results to the full-band (0.5--10\,keV) {\em Chandra} X-UDS catalogue described in Section \ref{xuds}. If we limit the UDS and the X-UDS to the same survey area (see Figure~\ref{fig:Data_field}, for reference), the samples are largely separate. Of the \varagn~variability detected active galaxies, \anyvarchandrasky~lie within the {\em Chandra} imaging region. The percentage of these that are dual-detected AGN (i.e.~X-ray detected and variable in the NIR) is only $\sim$\AGNoverlap~(\varxrayagn\,/\,\anyvarchandrasky). This represents $\sim$15 per cent of the X-ray detected sample (\varxrayagn\,/\,\xrayagn).

\begin{figure*}
    \centering
    \includegraphics[width=0.8\textwidth]{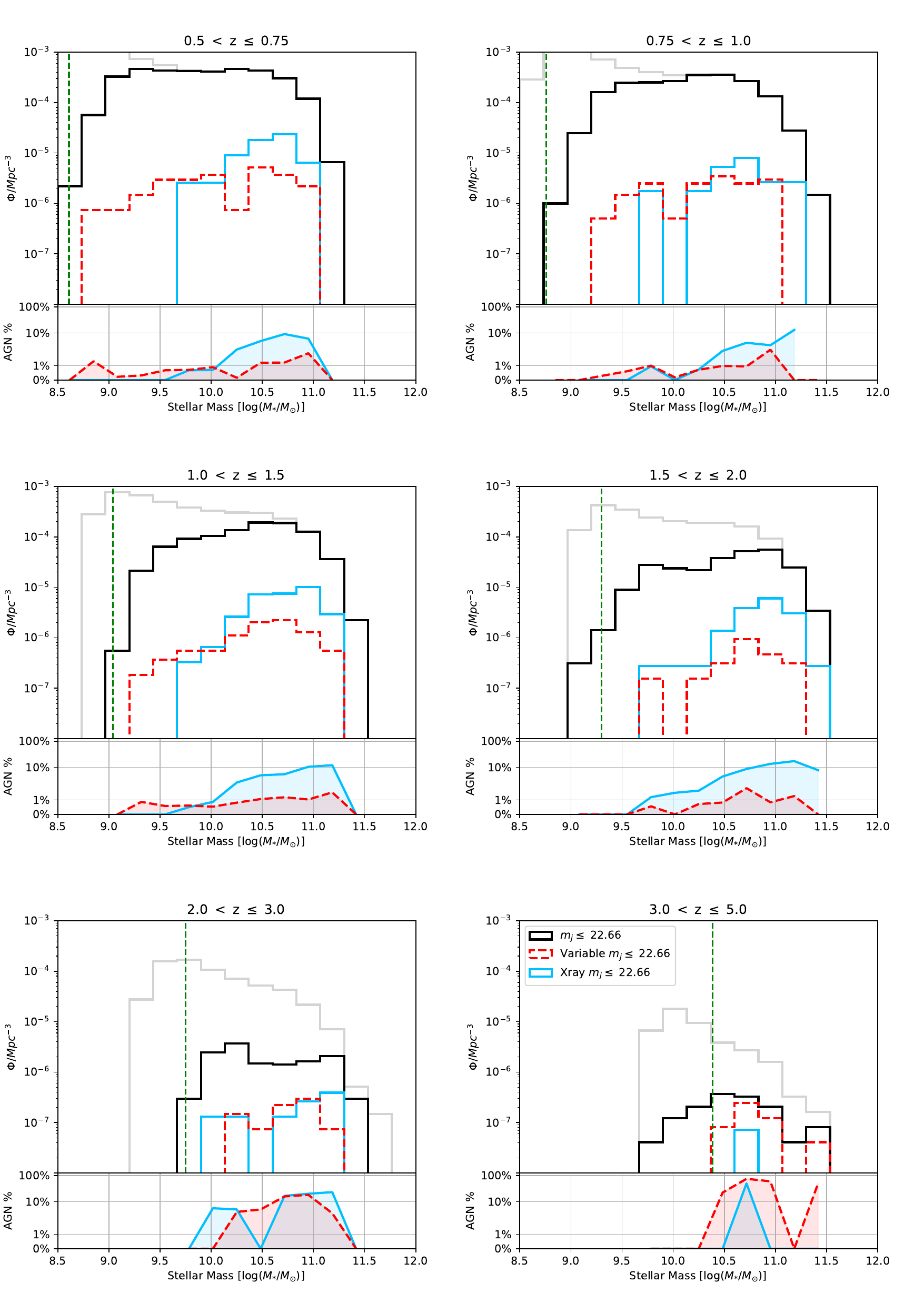}
    \caption{Stellar mass functions as a function of redshift. The grey histogram represents the mass distribution for all galaxies within the field except for quasar-like AGN, which have been removed. The black histogram shows the same but for galaxies with $J$-band apparent magnitudes $m_{J} \leq$ \jbandmag~which ensures that a variability of 20\% or higher will be consistently detectable. The dashed red histogram is as the black but shows only variability detected active galaxies.
    The solid blue histogram shows X-ray detected active galaxies, again with candidate quasars removed. X-ray detections were calculated using galaxies from the X-UDS imaging region and scaled to match the variability detected galaxies. The corresponding coloured distributions below each stellar mass function shows the percentage of AGN for that redshift bin as a function of stellar mass. The 90\% mass completeness limit is denoted by the vertical dashed green line.}
    \label{fig:massfunc}
    %/data/Var/Code/lum_func/Combined_mass_func1.py
\end{figure*} 

\begin{figure*}
    \centering
    \includegraphics[width=0.8\linewidth]{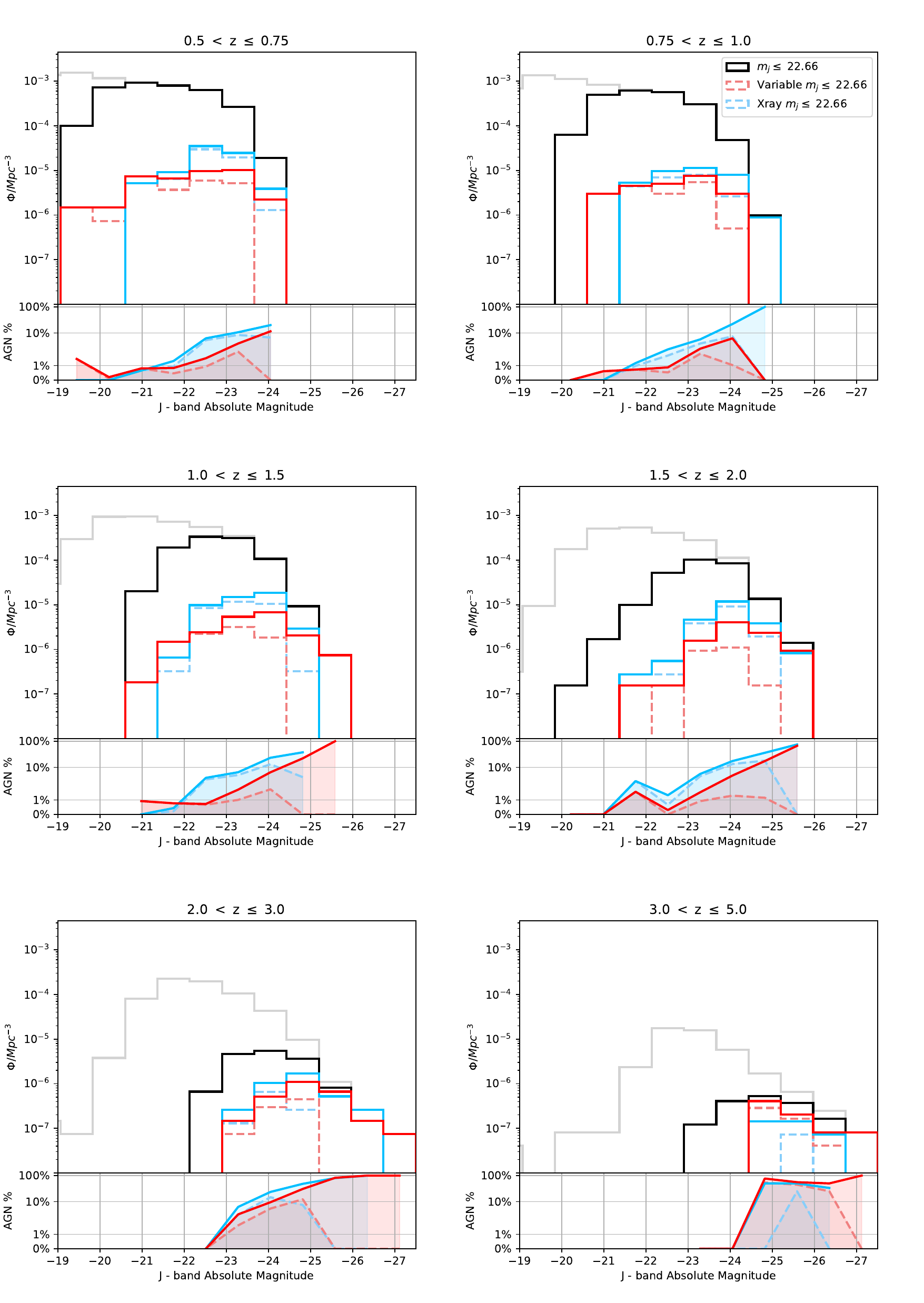}
    \caption{Luminosity functions as a function of redshift. The grey histogram represents the luminosity distribution for all galaxies within the field, while the black histogram shows the same but for galaxies with $J$-band apparent magnitudes $m_{J} \leq$ \jbandmag. The solid red and blue histograms are as the black but only shows variability and X-ray detected active galaxies, respectively. Lighter coloured, dashed histograms show the active galaxy luminosity functions with quasars removed. X-ray detections were calculated using galaxies from the X-UDS imaging region and scaled to match the variability detected galaxies. The corresponding coloured distributions below each luminosity function shows the percentage of AGN for that redshift bin as a function of decreasing $J$-band absolute magnitude.}
    \label{fig:lumfunc}
    %/data/Var/Code/lum_func/combined_lum_func3.py
\end{figure*}

One possible explanation for this result is that NIR-variable AGN have X-ray emission that is too faint to be detected at the depth of the X-UDS
survey. X-ray emission from AGN has long been known to correlate with optical/UV emission (e.g., \citealt{tananbaum_x-ray_1979}, \citealt{avni_x-ray_1986}). Therefore, to address this issue, we compare the observed X-ray--to-optical luminosity ratios to determine if the variability-detected AGN are unusually X-ray quiet (for galaxies that reside within the X-UDS survey region). In Figure~\ref{fig:lx_lo},  we compare the monochromatic luminosities at 2\,keV  and 2500$\,\text{\r{AA}}$, compared with the expected ratios assuming point-to-point spectral slopes with $\alpha_{ox}=1$ and $\alpha_{ox}=2$, corresponding to the typical range observed for quasars (e.g., \citealt{steffen_x-ray--optical_2006}). We note, however, that in our work we make no corrections for host galaxy contributions, dust reddening, or X-ray absorption. We are comparing the observed X-ray to optical ratios, to determine if the objects selected by variability show similar ratios to those selected with deep Chandra observations. Here, monochromatic luminosities at 2500$\,\text{\r{AA}}$ are derived from the observed optical/UV SEDs, given the source redshift, interpolating the flux between the nearest observed wavebands. The monochromatic X-ray luminosities are determined from the 0.5-10\,keV X-ray flux, assuming an X-ray photon index $\Gamma=1.9$ for all sources, as used in \citet{elmer_long-term_2020}. X-ray upper limits from \citet{kocevski_x-uds_2018} are used for X-ray non-detections. Overall we find that many variable active galaxies have similar optical luminosities to X-ray bright active galaxies, but the majority have only upper limits in our X-ray imaging. Based on the data available to us, the variable AGN appear systematically fainter in the X-rays.

One possible explanation is that the variable AGN are more obscured in the X-ray waveband, but deeper X-ray data will be needed to test that hypothesis. Overall, the implication is that near-infrared variability  appears to detect a set of active galaxies that are largely missed by X-ray surveys at these depths.

\subsection{Active Galaxy Properties}
\subsubsection{Stellar mass distributions and quasar contamination}
\label{sec: active gal properties}
In addition to there being limited overlap in AGN selected using NIR variability and X-ray detection, we find a difference in the stellar mass ranges of active galaxies selected via X-ray detection and NIR variability. Examining the mass distribution with redshift in Figure \ref{fig:mass_vs_z}, X-ray emission probes AGN in high-mass galaxies across all redshifts considered, with a majority of X-ray selected active galaxies having stellar masses $\geq 10^{10} \rm\,M_{\odot}$. This is in contrast to the active galaxies selected through variability, which show no obvious bias towards high-mass galaxies.  To illustrate these differences, 
in Figure \ref{fig:mass_hist} we plot the distribution of the stellar mass of the active galaxies, separated by detection method.
We find a clear distinction between any X-ray detected AGN, which are much more likely to exist in high-mass galaxies, and NIR variability detected AGN, which show a much more extended mass distribution.

One caveat to these results, however, is the potential for the contamination
by non-stellar light, which may bias the stellar masses for the brightest AGN.  To investigate this issue, we plot the measured stellar mass against the $J$-band absolute magnitude for all galaxies in the field in Figure \ref{fig:mass_mag}. 
Here we find the active galaxies to broadly exist in two regimes: one where the measured magnitude of the galaxy is much brighter than inactive galaxies of a similar mass, and one where the measured magnitude of the galaxy is similar to inactive galaxies of a similar mass. From this bimodality, we can assume that the observed light from the active galaxies that lie within the locus formed by the inactive galaxies are dominated by the hosts' stellar light, and as such the measured host galaxy parameters are minimally impacted by the presence of an AGN. On the other hand, active galaxies that are much brighter than their inactive counterparts of similar mass likely have emission that is dominated by the AGN, suggesting the presence of a quasar.  

To test if these abnormally bright sources are quasars, we investigate their morphology using the $K$-band stellarity index from \textsc{SExtractor}, which (after removing galactic stars) provides a good indication that an AGN is dominating the emission from a source. From this we find that a majority of the bright sources appear point-like, which suggests the AGN emission is outshining that of the host galaxy. 

These empirical trends are likely to be caused by a combination of factors. In particular, the stellar masses are obtained by fitting stellar population models, without an AGN component. If a Type-1 quasar dominates the SED, the best fit is likely to be an extremely young stellar population, yielding a very low mass-to-light ratio. As a secondary issue, the variable AGN were identified in the observed $J$ and $K$ bands, while observations at longer and shorter wavelengths were taken at different epochs. Therefore, as a selection effect, we might expect AGN to be brighter than average in the $J$ and $K$ bands, which would enhance their near-infrared luminosity relative to the stellar mass determined from the 12-band SED.

Furthermore, to investigate if a quasar is present within the identified bright, point-like sources, in Figure \ref{fig:mass_mag_lowsig} we remove the objects with the largest variability amplitudes, on the assumption that the NIR light in such objects must be dominated by the AGN. Figure \ref{fig:mass_mag_lowsig} shows the stellar mass vs. magnitude for galaxies with redshift $1<z\leq1.5$ both before and after requiring galaxies to have an intrinsic  variability amplitude of $\sigma_{\rm Q}\leq0.1$. Here we find that applying this limit primarily removes the point-like sources that are much brighter than inactive galaxies of a similar mass.  In the case of objects detected via their X-ray emission, imposing this limit does not have a significant effect on the active galaxies that lie within the locus formed by inactive galaxies, but largely removes the bright, point like sources. It is therefore likely that the bright point-like active galaxies are indeed quasars, where the powerful AGN is driving significant variability and dominating the NIR light.

\subsubsection{Stellar mass functions}
\label{sec: stellar_mass_func}

To fairly compare the host properties of the active galaxies, we remove the quasars from the distributions to avoid significant amounts of non-stellar light impacting measurements. To do this we generate a contour that encompasses 95\% of the inactive galaxies in each bin seen in Figure \ref{fig:mass_mag}, and we require the active galaxies in corresponding bins to lie within the contour formed by the inactive galaxies. Applying this cut to the data removes 75 (12\%) of the variable AGN from the sample, 35 (5\%) of the X-ray AGN and 55 (51\%) of the dual-detected AGN. Using the remaining galaxies, we generate the stellar mass functions shown in Figure \ref{fig:massfunc}. Here the grey line shows the mass function for all galaxies in the data and the solid black line shows the population of galaxies with $m_{J} \leq$ \jbandmag. As discussed in Section \ref{sec: var detect lim}, this magnitude limit ensures that a variability of 20\% or higher will be consistently detectable for all galaxies.

With the quasars removed, we find that the fraction of galaxies with X-ray detected AGN increases with increasing stellar mass of the host galaxy, whereas the detection rate of NIR variable AGN is largely flat, remaining on the order of $\sim1\%$ of galaxies regardless of the stellar mass of the host. This difference is consistent up to $z \approx 2$, where enough galaxies above the mass completeness limit are present to draw reasonable conclusions.  We conclude that, across a broad redshift range, NIR variability allows for the identification of AGN in much lower mass host galaxies compared to X-ray selection.

\vspace{-0.25cm}
\subsubsection{Luminosity functions}
\label{sec: luminosity_function}

In this section we investigate the luminosity functions of objects derived from the two methods of AGN detection, without applying any quasar filtering. This will provide a check on the influence of AGN light on the derived properties of these objects, which will be particularly significant in future situations where we may lack the range of spectral data and spatial resolution required to filter out quasars. In Figure~\ref{fig:lumfunc}, we plot the $J$-band absolute magnitude as a function of redshift for the detected active galaxies. 

Here, we find that both methods detect much more similar distributions of objects, with the fraction of AGN increasing with luminosity whichever technique is used.  This increase reflects the fact that the luminosity functions are strongly influenced by the contribution of AGN light, with brighter AGN more likely to be detected by either method.  Without the auxiliary information we have employed in this work, it is not possible to reliably disentangle host galaxy and AGN properties from such data.

\vspace{-0.6cm}

\section{Conclusions}
\label{sec:conclusion}

In this study, we use the long-term near-infrared variability of objects seen in the UKIDSS Ultra Deep Survey as an alternative approach to finding active galactic nuclei. We then compare the sample of AGN revealed in this way to those found using a more conventional X-ray-based method in the same field.

We find that the variability approach detects a distinct population of AGN, with an overlap of only \AGNoverlap\ between the two samples. An analysis of the X-ray-to-optical luminosity ratio finds the IR-variable AGN appear relatively X-ray quiet. Whether these AGN are intrinsically X-ray faint or heavily obscured is unclear, and will be the subject of future study.

After excluding quasar-like objects, whose AGN brightness means that we cannot reliably measure properties of the underlying host galaxy, we find that IR variability systematically detects AGN in galaxies with lower stellar masses than X-ray-detected objects. Comparing their mass functions, we find that IR-variable AGN are hosted in $\sim 1\%$ of galaxies of any mass, whereas the percentage of X-ray AGN detected strongly increases with increasing stellar mass of the host galaxy. This distinction seems to be independent of redshift out to the $z \sim 2$ probed by these data.  One plausible explanation for the difference would be related to extinction: if low-mass hosts tend to systematically contain more heavily obscured nuclei, that might explain why AGN in such galaxies are more readily detected by their infrared variability than by their extincted X-ray emission. In future work we will explore the dependence of variability on redshift and hence on rest-frame waveband, to explore the implications for the origin of the AGN emission.

In seeking such physical explanations, we need to be able to reliably to separate AGN and host properties. The significance of the contribution of AGN light to distorting our understanding of their host galaxies is underlined by the rather different results we obtain by simply studying their luminosity functions, in which a higher fraction of AGN is found in more luminous objects, whichever method is used to identify them. This distinction arises from the fact that the luminosity function, particularly when including quasar-like objects, reflects the properties of the AGN as well as their hosts.  Clearly, some care is needed in interpreting such composite data, and additional spectral and spatial information is vital.

These results indicate that multiple approaches to AGN detection are required to obtain a complete census of such objects. With upcoming deep, wide-field surveys such as EUCLID and LSST, which will be obtained over extended periods, long-term variability offers an important approach to identifying AGN missed by other methods, which is key to a more detailed understanding of the prevalence of these objects and their role in galaxy evolution.

\section*{Acknowledgements}

We wish to recognise and acknowledge that the mountain of Mauna a W\textoverline{a}kea is sacred to the indigenous Hawaiian community and has a very significant role in their culture. We are extremely grateful for the opportunity to conduct observations from this mountain. We also wish to thank the staff at UKIRT for their ongoing efforts in ensuring the success of the UDS project. This work is based in part on observations from ESO telescopes at the Paranal Observatory (programmes 180.A-0776, 094.A-0410, and 194.A2003).

%%%%%%%%%%%%%%%%%%%%%%%%%%%%%%%%%%%%%%%%%%%%%%%%%%
\section*{Data Availability}

The raw infrared imaging data used in this paper can be obtained from the WFCAM Science Archive (http://wsa.roe.ac.uk). Further processed data and catalogues will be made available from the UDS web page (https://www.nottingham.ac.uk/astronomy/UDS/). A public release of the processed data is in preparation (Almaini et al., in prep) and is available from the author on request (omar.almaini@nottingham.ac.uk).


\vspace{-0.5cm}

\bibliographystyle{mnras}
\bibliography{references} % if your bibtex file is called example.bib

%%%%%%%%%%%%%%%%%%%%%%%%%%%%%%%%%%%%%%%%%%%%%%%%%%

%%%%%%%%%%%%%%%%% APPENDICES %%%%%%%%%%%%%%%%%%%%%

% \appendix

% \section{Some extra material}

% If you want to present additional material which would interrupt the flow of the main paper,
% it can be placed in an Appendix which appears after the list of references.

%%%%%%%%%%%%%%%%%%%%%%%%%%%%%%%%%%%%%%%%%%%%%%%%%%

% Don't change these lines
\bsp	% typesetting comment
\label{lastpage}
\end{document}